\documentclass[9pt,preprint,numbers]{sigplanconf}

\usepackage{amsmath}

\usepackage{subfig}
\usepackage{multirow}
\usepackage{array}

\newcommand{\XPTO}{NG2C}

\usepackage{mathtools}
\usepackage{listings}
\usepackage{color}
\usepackage{algorithm}
\usepackage[noend]{algpseudocode}

\definecolor{dkgreen}{rgb}{0,0.6,0}
\definecolor{gray}{rgb}{0.5,0.5,0.5}
\definecolor{mauve}{rgb}{0.58,0,0.82}

\begin{document}

\setlength{\pdfpageheight}{\paperheight}
\setlength{\pdfpagewidth}{\paperwidth}

\conferenceinfo{ISMM '17}{June 18--23, 2017, Barcelona, Spain}
\copyrightyear{20yy}
\copyrightdata{978-1-nnnn-nnnn-n/yy/mm}
\copyrightdoi{nnnnnnn.nnnnnnn}



\title{\XPTO: Pretenuring N-Generational GC for HotSpot Big Data Applications}

\authorinfo{Rodrigo Bruno}
           {INESC-ID / Instituto Superior T\'{e}cnico, University of Lisbon}
           {rodrigo.bruno@tecnico.ulisboa.pt}
\authorinfo{Lu\'is Oliveira}
           {Feedzai}
           {luis.oliveira@feedzai.com}
\authorinfo{Paulo Ferreira}
           {INESC-ID / Instituto Superior T\'{e}cnico, University of Lisbon}
           {paulo.ferreira@inesc-id.pt}

\maketitle


\begin{abstract}
Big Data applications suffer from unpredictable and unacceptably
high pause times due to Garbage Collection (GC). This is the 
case in latency-sensitive applications such as on-line 
credit-card fraud detection, graph-based computing for analysis 
on social networks, etc. 
Such pauses compromise latency requirements of the whole
application stack and result from applications' 
aggressive buffering/caching of data, 
exposing an ill-suited GC design, which assumes that most
objects will die young and does not consider that applications 
hold large amounts of middle-lived data in memory.

To avoid such pauses, we propose \XPTO,
a new GC algorithm that combines pretenuring with an N-Generational
heap. By being able to allocate objects into different generations,
\XPTO~is able to group objects with similar lifetime profiles in the
same generation. By allocating objects with similar lifetime profiles
close to each other, i.e. in the same generation, we avoid
object promotion (copying between generations) and heap fragmentation 
(which leads to heap compactions) both responsible for most of the duration of 
HotSpot GC pause times.

\XPTO~is implemented for the OpenJDK 8 HotSpot Java
Virtual Machine, as an extension of the Garbage First GC. 
We evaluate \XPTO~using Cassandra,
Lucene, and GraphChi with three different GCs:
Garbage First (G1), Concurrent Mark Sweep (CMS), and \XPTO.
Results show that \XPTO~decreases the worst observable 
GC pause time by up to 94.8\% for Cassandra, 85.0\% for 
Lucene and 96.45\% for GraphChi, when compared to current 
collectors (G1 and CMS). In addition, \XPTO~has no negative 
impact on application throughput or memory usage.

\end{abstract}



\section{Introduction}
Big Data applications are now part of the application stack present in
most (if not all) large-scale systems. These applications are expected
to work with high volumes of information efficiently and often run on top 
of platforms such as Cassandra~\cite{lakshman-2010}, 
Lucene~\cite{mccandless-2010}, GraphChi~\cite{kyrola-2012}, etc. 
This is the case of latency-sensitive applications such as on-line credit-card 
fraud detection, graph-based computing for analysis on social networks 
or the web graph, etc. 

To achieve good performance, developers often resort to optimization techniques
to boost performance such as caching \cite{power-2010,shinnar-2012,zhang-2015}. 
Caching is used to keep (in memory): i) the working set or intermediate results 
\cite{zaharia-2012} (this is a common practice, for example, 
in graph processing systems such as GraphChi \cite{kyrola-2012} and Spark 
\cite{zaharia-2010}), or ii) consolidate writes in a database (for example,
in-memory tables in Cassandra \cite{lakshman-2010}).
With caching, developers avoid costly operations such as recomputing intermediate 
values (in the case of GraphChi and Spark) or writing records to disk 
(in the case of Cassandra), among others. However, while keeping more data in memory 
helps reducing the latency for data requests, it puts more pressure on the 
Garbage Collection (GC) which results in long pauses of the application
\cite{bu-2013,ousterhout-2015}.

GC pauses are unpredictable (from the application's perspective) and
can stop the application for an unacceptably high amount of time
leading to broken Service Level Agreements (SLAs)~\cite{dean-2013,ongaro-2011,rumble-2011}. 
The is even worse if the application stack contains multiple managed 
heaps. If only one GC engages in a long GC pause, the whole stack is compromised
and the SLA is broken.

By analyzing applications running on the HotSpot JVM, it is possible to conclude
that the duration of
GC pauses is dominated by the number 
and size of objects to copy in memory during a GC (this problem is also described in 
Gidra \cite{gidra-2013,gidra-2011}). Such copies can be triggered by object promotion
or by a heap compaction (to reduce fragmentation). The problem with copying is that it 
is bound to the available hardware memory bandwidth. Increasing the capacity of nodes 
(e.g., the number of cores) will not reduce neither the number nor the duration of GC 
pauses. 


Therefore, the widely accepted weak generational hypothesis stating that most objects die young \cite{jones-2006,jones-2008,ungar:1984} (which is a 
fundamental design rule 
for current HotSpot GCs) is not suited for many Big Data applications. As a matter of fact, 
such applications try to fit their working set in memory, leading to a high number 
of objects that live for a long period of time (from the GC perspective). 
This mismatch between the objects' real lifetime and the GC assumption that 
most objects die young has serious consequences for HotSpot applications with 
high memory utilization and tight response time targets (i.e., SLAs to cope with).

Avoiding object copying within the heap cannot (or is extremely hard to)
be attained by tweaking the heap or GC parameters; thus, many
application developers end up to (almost) reverse engineering the GC to
understand how to avoid costly GC pauses in their applications \cite{lengauer-2014} 
(with obvious software development productivity drawbacks). Another
important difficulty is that a particular GC/heap configuration will only
work for a specific environment (number of cores, size of memory,
etc.), making a particular configuration not replicable.

Previous works
\cite{cheng:1998,blackburn:2001,marion:2007,harris:2000,jump:2004,clifford:2015}
used pretenuring to reduce the amount of object copying, however, with limited
success \cite{jones-2016}. The main problem with these solutions is that
simply pretenuring objects to a single older space (old generation) leads to
heap fragmentation since objects with different lifetime profiles are promoted
to the same space. \XPTO, opposed to previous pretenuring works, uses an N-Generational
heap that groups objects with similar lifetime profiles in the same generation.
This reduces object promotion and heap fragmentation. See Section \ref{sec:rel} for more details.


In order to use \XPTO, object allocation sites must be annotated to indicate in which
generation the object should be allocated. To free the
developer from the burden of understanding the objects' lifetime profiles, we developed 
a profiler tool, Object Lifetime Recorder (described in Section \ref{sec:tools}) that 
profiles the application and outputs where and how the code should be changed in order 
to take full advantage of \XPTO~(note that the application only needs to be profiled
once).



\XPTO~is implemented for the OpenJDK 8 HotSpot Java Virtual Machine (JVM) as
an extension of the next OpenJDK by-default GC, Garbage First (G1). 
Results are very encouraging as we are able to achieve our goal: avoid costly 
object copying (which occurs during object promotion and compactions). 
\XPTO~decreases the highest observable GC pauses by 
up to 94.8\% for Cassandra, 85.0\% for Lucene and 96.45\% for GraphChi, when 
compared to current collectors: G1 and CMS (Concurrent Mark Sweep). 
In addition, application throughput and memory usage are not negatively 
affected by using \XPTO. 

To sum up, our main contributions are: i) the identification of the
scalability problem of current HotSpot GCs which rely on object copying, 
ii) a new GC algorithm, \XPTO, that combines an arbitrary number of
generations with pretenuring to avoid object copying and heap fragmentation, 
iii) an implementation of \XPTO~and the Object Lifetime Recorder (OLR) profiler 
tool on a production JVM, OpenJDK 8 HotSpot, and iv) the evaluation of the performance 
benefits of using \XPTO~on Big Data platforms (Cassandra, Lucene, GraphChi) using 
workloads and data sets based on real system utilization.

\section{Related Work}
\label{sec:rel}
\XPTO~proposes N-Generational pretenuring, where objects are pretenured
into different generations, according to their lifetime profile. By allocating
objects in different generations, per-lifetime profile, 
\XPTO~reduces object promotion and heap fragmentation. Since this work combines 
several well studied ideas, we dedicate this section to explaining how our 
research relates to previous work.

\subsubsection*{Generational Collectors}
Segregating objects by age has been studied for a long time \cite{jones-2016} 
as a way to take advantage of the weak generational hypothesis 
\cite{ungar:1984}. By promoting objects that survive a number of collections 
into older generations, the collector can concentrate on collecting younger
generations more often (since these are more likely to contain more dead 
objects) \cite{lieberman:1983}. The use of multiple generations (compared to 
using a single generation) has been shown to reduce application pauses 
\cite{seligmann-1995,hudson-1997,blackburn-2002,marlow:2008}.

Opposed to previous works such as the Beltway framework \cite{blackburn-2002} and
the Mature Object Space collector \cite{hudson-1992}, \XPTO~
does not promote/copy 
objects gradually through older generations since this only generates more object
copying (which we are trying to avoid). Instead, objects are pretenured based on 
the object lifetime profile.

\subsubsection*{Pretenuring and Object Demographics}
Pretenuring is also a well studied technique 
\cite{cheng:1998,blackburn:2001,marion:2007,harris:2000,jump:2004,clifford:2015}. 
It consists on allocating objects (that are known to live for a long time) 
directly in older generations. By doing this, the overhead associated to object
promotion is avoided. The key problem to pretenuring lies on how to 
estimate the lifetime profile of an object (analyzed next).

To the best of our knowledge, no pretenuring algorithm has been combined
with an N-Generational collector. If a collector with predefined number of 
generations is used (for example, two generations), pretenuring
does not solve heap fragmentation, as middle-long lived objects with different 
lifetime profiles might be pretenured to the same heap location. To solve this 
problem \XPTO~combines pretenuring with multiple generations. By being able to 
pretenure into an arbitrary number (defined at runtime) of generations,
\XPTO~avoids fragmentation. Thus, by combining these two techniques, object copying 
(which results from object promotion and compaction) is greatly reduced.

Pretenuring is tightly coupled with object lifetime profiling, which is used
to extract object lifetime estimations, used to guide 
pretenuring. Extracting objects demographic information can either
be performed dynamically \cite{harris:2000,jump:2004,clifford:2015} 
or statically \cite{cheng:1998,blackburn:2001,marion:2007}. Profiling
information can come from stack analysis \cite{cheng:1998,clifford:2015},
connectivity graphs \cite{guyer:2004} and can also include other program's
traces \cite{marion:2007}.

Our proposed profiler, (presented in Section \ref{sec:tools}) builds upon previous
works \cite{cheng:1998,blackburn:2001,marion:2007} by resorting to stack analyzes. 
However, opposed to previous works, it accurately estimates
in which generation an object should be allocated in. In other words, our profiler 
answers the question of how long will the object probably live while previous
profilers only tell us if the object will probably live enough to be considered 
old (this information is not sufficient to take advantage of an 
N-Generational heap).

\subsubsection*{Region-based Garbage Collection}
The hypothesis that many objects, allocated in the same scope, share the same 
faith, i.e., have similar lifetimes has also been leveraged by many 
region-based memory management algorithms 
\cite{nguyen-2015,gog-2015,kowshik-2002,hicks-2004,hallenberg-2002,grossman-2002,gay-2001,davidson-1998,boyapati-2003,beebee-2001,aiken-1995,gay-2000}.
In such algorithms, objects with similar lifetime profiles are allocated in 
scope-based regions, which are deallocated as a whole when objects 
inside these regions are no longer reachable. However, existing region-based 
algorithms either require sophisticated static analysis
\cite{aiken-1995,boyapati-2003,beebee-2001,gay-2000,gay-2001,grossman-2002} 
(which does not scale to large systems), heavily rely on manual code 
refactoring \cite{nguyen-2015,gog-2015} (to guarantee that objects in the same 
region die approximately at the same time), or support only simple programming
models \cite{lu-2016,nguyen-2015,nguyen-2016} (such as parallel bag of tasks).

Other region-based collectors \cite{doligez:1993} use thread-local 
allocation regions to allocate objects. This approach also does
not support more complex models where most large data structures
can be maintained by multiple threads (for example, Cassandra's in-memory tables).

\XPTO~can also be seen/used as a region-based collector, in which generations
can be used as regions. However, opposed to typical scope-based regions,
\XPTO~supports more complex programs such as storage platforms with minimal
code changes.

\subsubsection*{Off-heap based Solutions}
There are some solutions based on off-heap memory 
\cite{mastrangelo-2015,nguyen-2015,lu-2016} 
(i.e., allocating memory for the application outside the GC-managed heap). 
While this is an effective approach to allocate and keep data out of the 
range of the GC (and therefore, reducing object copying), it has several 
important drawbacks: 
i) off-heap data needs to be serialized to be saved in off-heap memory, and 
de-serialized before being used by the application 
(this obviously has performance overheads); ii) off-heap memory must be
explicitly collected by the application developer (which is error prone
\cite{caballero-2012,cowan-2000} and completely ignores the advantages of 
running inside a memory managed environment); iii) the application must always
have objects identifying the data stored in off-heap (these so called
header objects are stored in the managed heap therefore stressing the GC). 
Furthermore, as shown in Section \ref{sec:evaluation}, \XPTO's approach 
outperforms off-heap memory.

To conclude our analysis of related work, to the best of our knowledge,
this work is the first to: i) combine N generations with pretenuring, and
ii) to show that it reduces object copying (coming from object promotion and compaction),
that affects many HotSpot Big Data applications, thus improving applications' performance.

Although \XPTO's observable benefit is the reduction of application pause times,
our contribution is orthogonal to other techniques used to implement low pause time collections
such as incremental (for example Immix \cite{blackburn:2008} and G1 \cite{detlefs-2004})
or concurrent compaction (for example C4 \cite{tene-2011}), and to real-time collectors
(for example the work in Fiji VM \cite{pizlo:2009} and the Metronome \cite{bacon:2003}
collector).

In fact, we envision that  
\XPTO~could be integrated with such algorithms to improve the collector's 
performance and reduce interference in application performance.



\section{Pretenuring N-Generational GC}

\begin{figure}[t]                                                  
\centering                                                                   
\includegraphics[keepaspectratio,width=.4\textwidth]{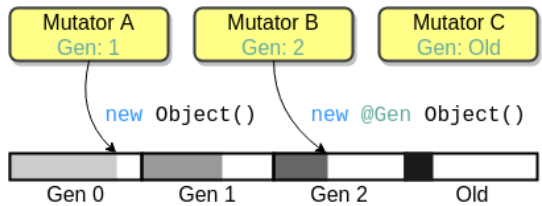}
\caption{Allocation of Objects in Different Generations}
\label{fig:ng2c-arch}
\end{figure}

In this section we provide a description of the proposed solution, that combines
an N-Generational heap layout with pretenuring. This algorithm extends the
G1\cite{detlefs-2004} collector, from which we reuse many mechanisms. In Section
\ref{sec:impl}, we describe how \XPTO~is implemented on top of G1.




\subsection{N-Generational Heap Layout}
\XPTO~builds upon generational collectors's \cite{appel-1989} idea but provides an arbitrary
number of generations. The concept of generation is used instead of local/private allocation
region because objects are grouped by estimated lifetime/age instead of being grouped by 
the allocating thread.

The heap is always created with two generations: \textit{Gen 0} and \textit{Old}. By default, 
all objects are allocated in
\textit{Gen 0}. Upon collection (more details in Section \ref{sec:collection}), 
live objects are promoted to the \textit{Old} generation. In other words, if no
new generations are created, \XPTO's heap layout is a 2-generational heap layout.

At run time, any number of extra generations might be created (see Section 
\ref{sec:using_ng2c} for more details). Objects can be allocated directly in
each of these generations and the percentage of heap usage for each
generation grows dynamically with the amount of allocated objects. 
With time, when objects become unreachable, the space previously allocated for
a specific generation becomes available for other generations to use
(more details in Section \ref{sec:collection}). In \XPTO, except for the \textit{Gen 0},
the amount of heap space assigned to each generation is dynamic, increasing or decreasing
as the amount of objects in that particular generation increases or decreases, respectively.
This is possible since each generation is not implemented as a single large block of memory, but instead, as a linked list of memory regions (more details in Section \ref{sec:impl}).

\begin{lstlisting}[float,floatplacement=H,label=lst:api,caption={\XPTO~API}]
// Methods added in class java.lang.System:
public static Generation newGeneration();
public static Generation getGeneration();
public static void setGeneration(Generation);
\end{lstlisting}

\subsection{N-Generational Pretenuring}
\label{sec:using_ng2c}
\XPTO~is designed to profit from information regarding objects' lifetime profiles 
(as described in Section \ref{sec:tools}, this information is provided by the OLR
profiler). Thus, \XPTO~co-locates objects with different lifetime profiles 
in different generations. 

Since applications might have multiple threads/mutators managing objects with
different lifetime profiles (e.g., processing separate jobs), each
thread must be able to allocate objects in different generations.

To efficiently support parallel allocation in multiple generations,
we bind each application thread into a specific generation using the
concept of current generation. The current generation indicates the generation 
where new objects, allocated with the {\tt@Gen} annotation\footnote{Starting from Java 8,
the {\tt new} operator can be annotated. We use this new feature to place a special
annotation that indicates that this object should go into the thread's current generation.}, 
will be allocated into. In practice, 
when a thread is created, its current generation is \textit{Gen 0}. If the 
thread decides to create a new generation, this will change the thread's current generation 
to the new one. It is also possible to get and set the thread current generation.

More specifically, the application code can use the following calls (see 
Listing \ref{lst:api}): 

\textbullet\hphantom{s}{\tt newGeneration}, creates a new generation and sets the
  current generation of the executing thread to the newly created
  generation;
  
\textbullet\hphantom{s} {\tt getGeneration} and {\tt setGeneration}, gets and sets
  (respectively) the current generation of the executing thread.

To allocate an object in the current generation, the {\tt new} operator must be
annotated with {\tt@Gen}. All allocation sites with no {\tt@Gen} will allocate
objects into \textit{Gen 0}  (see Figure \ref{fig:ng2c-arch}).

The code example in Listing \ref{lst:job} resembles a very simplified version
of graph processing systems (e.g., GraphChi). It shows a method that runs 
several tasks in parallel threads. Each thread starts by calling 
{\tt newGeneration}, to create a new generation. Then, while the task 
is not finished, all allocated objects using the {\tt @Gen} annotation will
be allocated in the new generation.

\begin{lstlisting}[float,floatplacement=H,label=lst:job,caption={Job Processing Code Sample}]
public void runTask() {
  Generation gen = System.newGeneration();
  while (running) {
    DataChunk data = new @Gen DataChunk();
    loadData(data);
    doComplexProcessing(data);}}
\end{lstlisting}

\begin{lstlisting}[float,floatplacement=H,label=lst:cache,caption={Data Buffer Code Sample}]
public class Buffer {
	byte[][] buffer;
	Generation gen;
	public Buffer() {
		gen = System.newGeneration();
  		buffer = new @Gen byte[N_ROWS][ROW_SIZE];}}
\end{lstlisting}

Listing \ref{lst:cache} shows a code example that resembles a very simplified
version of memory buffers in storage systems such as Cassandra; it shows how to 
use \XPTO~to allocate a large data structure (e.g., a buffer to consolidate database 
writes or intermediate data) while avoiding object copying. The constructor creates 
a new generation in which the buffer is allocated (using the {\tt @Gen} annotation). 


\makeatletter
\def\BState{\State\hskip-\ALG@thistlm}
\makeatother

\begin{algorithm}[!t]
\caption{Memory Allocation - Object Allocation}
\label{alg:mem_alloc_object}
\begin{algorithmic}[1]
\Procedure{Object Allocation}{}

\State $\textit{size} \gets \text{size of object to allocate}$
\State $\textit{klass} \gets \text{class of object to allocate}$
\State $\textit{isArray} \gets \text{object is of array type?}$
\State $\textit{gen} \gets \text{current thread generation}$
\State $\textit{isGen} \gets {\tt new} \text{ operator annotated with } {\tt @Gen} \text{?}$

\If {$isGen$} 
\State $\textit{tlab} \gets \text{TLAB used for generation } \textit{gen}$
\Else
\State $\textit{tlab} \gets \text{TLAB used for} \textit{ Gen 0 }$
\EndIf

\If {$isArray$}
\State \textbf{goto} \emph{slow path}
\EndIf

\If {$end(tlab) - top(tlab) >= size$}
\State $\textit{object} \gets \textit{init(klass, top(tlab))}$
\State $\textit{bumpTop(tlab, size)}$
\State $\Return \textit{ object}$
\EndIf

\BState \emph{slow path}:

\If {$size >= size(tlab)/8$}
\State $\Return \Call{Alloc In Region}{klass, size}$
\Else
\State $\Return \Call{Alloc In TLAB}{klass, size}$
\EndIf
\EndProcedure
\end{algorithmic}
\end{algorithm}

\begin{algorithm}[!t]
\caption{Memory Allocation - Allocation in Region}
\label{alg:mem_alloc_region}
\begin{algorithmic}[1]
\Procedure{Alloc in Region}{klass, size}

\State $\textit{gen} \gets \text{current thread generation}$
\State $\textit{isGen} \gets {\tt new} \text{ operator annotated with } {\tt @Gen} \text{?}$

\If {isGen} 
\State $\textit{region} \gets \textit{ gen } \text{alloc region}$
\Else
\State $\textit{region} \gets \textit{ Gen 0 } \text{alloc region}$
\EndIf

\If {$end(region) - top(region) >= size$}
\State $\textit{object} \gets \textit{init(klass, top(region))}$
\State $\textit{bumpTop(region, size)}$
\State $\Return \textit{ object}$
\EndIf

\If {isGen}
\State $\textit{region} \gets \text{new} \textit{ gen } \text{alloc region}$
\Else
\State $\textit{region} \gets \text{new} \textit{ Gen 0 } \text{alloc region}$
\EndIf

\If {$\textit{region} \text{ not null}$}
\State $\textit{object} \gets \textit{init(klass, top(region))}$
\State $\textit{bumpTop(region, size)}$
\State $\Return \textit{ object}$
\Else
\State trigger GC and retry allocation
\EndIf
\EndProcedure
\end{algorithmic}
\end{algorithm}

\subsection{N-Generational Memory Allocation}
\label{sec:allocation}
\XPTO~allows each thread to allocate objects in any generation. This is 
fundamentally different from current HotSpot's allocation strategy which assumes that
all newly allocated objects are placed in the young generation. 
Hence, in order to support N-Generational object allocation, we extended the
allocation algorithm to be able to allocate objects in multiple generations.

In \XPTO, object allocation is separated in two paths: 
i) fast allocation path, using a Thread Local Allocation Buffer 
(TLAB)\footnote{A TLAB is a Thread Local Allocation Buffer, i.e., a private buffer
where the thread can allocate memory without having to synchronize with other 
threads.}, and ii) slow allocation path (array or very large object allocation). 

Allocations through the {\tt slow path} are handled in one of two ways: inside a
TLAB (if there is enough free space), or directly in the current 
Allocation Region (AR)\footnote{An Allocation Region is used to satisfy
allocation requests for large objects and also for allocating TLABs. Whenever
an AR is full, a new one is selected form the list of available regions.} 
(outside a TLAB).
Note that for each generation, there is one AR.

The high level algorithm is depicted in Algorithms \ref{alg:mem_alloc_object} and 
\ref{alg:mem_alloc_region}.
For the sake of simplicity, and without loss of generality, 
we keep the algorithm description to the minimum, only keeping the most 
important steps.

A call to {\tt Object Allocation} starts an object allocation. 
If the allocation is marked with {\tt @Gen}, the allocation takes place 
in the current generation which is available from the executing thread 
state (otherwise the object is allocated in \textit{Gen 0}). 
Arrays automatically fall into the {\tt slowpath} while regular 
objects are promptly allocated from the TLAB unless there is not enough space.

Large object allocation (objects larger than a specific fraction of the TLAB size) 
goes directly to the current AR of the current generation (or \textit{Gen 0} 
if the allocation is not annotated). If the region has enough
free space to satisfy the allocation, the object is allocated.
Otherwise, a new region is requested from the available regions' list within
the heap. If no memory is available for a new region, a GC
is triggered followed by an allocation retry. If a GC
is not able to free enough memory, an error is reported to the application.

The pseudocode for allocations in TLABs is not shown because of space 
restrictions. Nevertheless, the code between lines 7 and 16 is already 
representative of how allocations inside a TLAB are conducted.

\subsection{N-Generational Memory Collection}
\label{sec:collection}

\begin{figure}[t]                                                  
\centering                                                                   
\includegraphics[keepaspectratio,width=.45\textwidth]{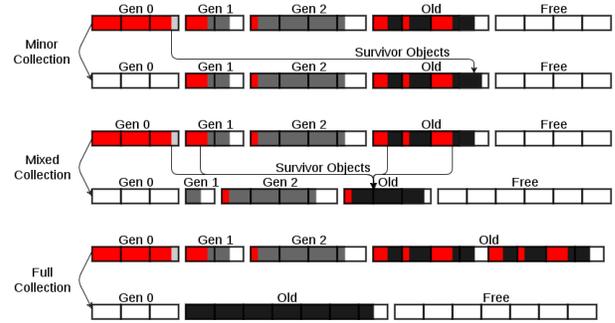}
\caption{Types of collections (red represents unreachable data)}
\label{fig:ng2c-collections}
\end{figure}

In \XPTO, three types of collections can take place (see Figure \ref{fig:ng2c-collections}):

\textbullet\hphantom{s}minor collection: triggered when \textit{Gen 0} has no space
left for allocating new objects. Collects the \textit{Gen 0}. Objects that survived 
a number of collections (more details in Section \ref{sec:impl}) are promoted to the
\textit{Old} generation;

\textbullet\hphantom{s}mixed collection: triggered when \textit{Gen 0} has no space
left for allocating new objects and the total heap usage is above a configurable threshold. Collects 
the \textit{Gen 0} plus other memory regions from multiple generations whose amount of 
live data is low (more details in Section \ref{sec:impl}). Survivor objects from any 
of the collected memory regions (excluding memory regions from the \textit{Old} generation),
are promoted to the \textit{Old} generation. Please note that, in a mixed collection,
although all the regions belonging to \textit{Gen 0} are collected, regions belonging to
other generations are only collected if the percentage of live data is below a configurable
threshold. In addition to collecting multiple regions from multiple generations, a mixed
collection also triggers a concurrent marking cycle (described next);

\textbullet\hphantom{s}full collection: triggered when the heap is nearly full. Collects the 
whole heap. In a full collection, all regions belonging to all generations are collected. 
All survivor objects are promoted to the \textit{Old} generation. 

Note that when all regions that compose a generation are collected, the generation is discarded.
If future allocations that target a specific generation that was previously discarded, the target 
generation is re-created before the first allocation is actually performed.

In a concurrent marking cycle (that starts during a mixed collection), the GC traverses the heap and 
marks live objects. As the name indicates, most of this process is done concurrently with the 
application. When the marking phase ends, the GC frees all regions containing only unreachable 
(i.e., unmarked) objects. For the regions that still contain reachable content, the 
GC saves some statistics (used for example in mixed collections) on how much memory can be 
reclaimed if a particular region is collected.

\subsection{Object Lifetime Recorder}
\label{sec:tools}

\begin{figure}[t]                                                  
\centering                                                                   
\includegraphics[keepaspectratio,width=.4\textwidth]{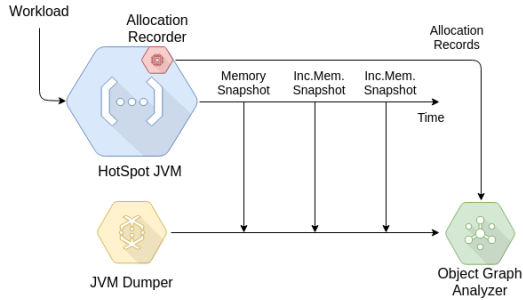}
\caption{Object Lifetime Recorder Profiler Architecture}
\label{fig:olr-arch}
\end{figure}

To enable developers to take full advantage of \XPTO, we developed the Object Lifetime Recorder 
(OLR) profiler, a HotSpot JVM profiler that records object allocation sites and lifetimes.
Using OLR, no developer's knowledge is required to change the code in order to take advantage
of \XPTO. 

OLR is composed by three components (see Figure \ref{fig:olr-arch}). First, the Allocation Recorder (implemented as
a Java Agent\footnote{A Java Agent is a small
pluggable component that can be attached to the JVM, being able to analyze its state.}) 
is used to: i) notify the JVM Dumper (second component, described next) when a collection 
finishes, and ii) record allocation sites. The second component, the JVM Dumper creates 
incremental heap dumps\footnote{A heap dump is a memory snapshot (taken while the
application is running) of the Java heap (where all the application objects reside). We create incremental dumps using CRIU, a Checkpoint/Restore tool for Linux processes.} (regarding
previous heap dumps, taken upon previous collections) whenever a collection finishes
(the JVM Dumper is notified by the Allocation Recorder when a heap dump should be taken). 
Compared to other heap dump tools, for example, with the {\tt jmap} tool, incremental dumps 
are smaller in size (as they contain only modified memory positions), thus leading to shorter 
application stop times for creating the heap dump.

The third component, the Object Graph Analyzer is used to process the allocation sites 
and heap dumps generated during the application profiling. Objects' metadata (allocation 
timestamp, collection timestamp, and allocation site) is loaded into memory and an object
graph is built. Then, the graph is processed in order to extract and present which objects
should belong in the same generation, and where these objects are allocated.

In practice, even an inexperienced developer can change the source code of an
application to take advantage of \XPTO. The developer only needs to run the application
using OLR's Allocation Recorder and run the JVM Dumper. When the application finishes 
(or after running for some time under the normal/target workload), the
developer launches the Object Graph Analyzer that outputs where and how the code should
be modified. With this information, even an inexperienced programmer can
change the code locations suggested by the OLR in minutes.

We measured a performance cost (throughput) of up to 4 times when running the profiler on 
the systems we use in the evaluation. However, note that the profiler only has to run once and 
that the code changes proposed by the profiler lead to significant performance improvements 
in production settings (as observed in Section \ref{sec:evaluation}). 
In other words, despite the fact that the profiler analyzes the application with reduced 
throughput, the captured allocation and collection patterns hold true in a production setting.

\section{Implementation}
\label{sec:impl}
\XPTO~is implemented on top of the Garbage First (G1) GC \cite{detlefs-2004}. G1 is the most recent 
and advanced GC algorithm available for the OpenJDK HotSpot JVM 8. In addition, 
G1 is the new default GC in the HotSpot JVM. \XPTO~builds upon G1, by adding 
approximately 2000 LOC. \XPTO~is integrated with G1 in the way that applications that do
not use the {\tt @Gen} annotation will run using the G1 collector (i.e., the code introduced by 
\XPTO~is never activated). This has the great benefit that all the effort put into 
developing G1 ensures that \XPTO~works with the same performance to G1 for all applications. 
For the rest of this section, we describe how we modified G1 for supporting 
N-Generational pretenuring.

By using G1's as our code base, we inherit many techniques that are already
well implemented and tested. In other words, we are using all the GC techniques 
already implemented in G1 (such as remembered sets management, safepoints, 
write barriers, and concurrent marking) to support \XPTO's implementation. 

G1 uses a heap divided in equally sized memory regions. It contains two 
generations, the young and the old. The young generation is divided into three
spaces \cite{ungar:1984} (\textit{Eden}, and two survivor spaces, \textit{To}, and
\textit{From}). In \XPTO, \textit{Gen 0} corresponds to the young generation in G1
(including the three spaces composing it). The old generation corresponds to \XPTO's
\textit{Old} generation. Additional generations are created by allocating regions
from the free memory regions list (also available in G1). The existing code in G1
looks at \XPTO's additional generations as part of G1's old generation. This means
that we reuse G1's write barrier and remembered set for inter-generational pointers.

\XPTO's inherits G1's collector algorithms without almost any change. Minor, mixed,
and full collections work in the exact same way in both \XPTO and G1. The only
modification is that, in \XPTO, the collector can promote objects from different
generations into the \textit{Old} generation, while in G1, the collector either only
promotes from young to old or compacts regions belonging to the old generation.

Most of the code introduced by \XPTO, lies in the object allocation path. In the
next sections we describe how the new allocation algorithm works and how the 
byte code interpreter and Just-In-Time compiler are adapted to work with it.

\subsection{Parallel Memory Allocation}
\label{sec:par_alloc}
Contention in memory allocation is a well-known problem \cite{jones-2016,gay-2000};
memory allocation must be synchronized between threads so that each memory 
block is used by a single thread. In G1 this is achieved by using Thread
Local Allocation Buffers (TLABs) and Allocation Regions (ARs). 
Therefore, whenever a thread needs to allocate some memory, it
allocates directly from its TLAB. 
If the TLAB is full, the thread must allocate memory from the current AR.
This allocation, however, will only occur after the thread acquires a lock on that 
AR. If the AR does not have enough available space, a new AR is allocated directly
from the list of free regions (this step requires even further locking to ensure that no
other thread is allocating another region).

In \XPTO, we extend the use of both TLABs and ARs to multiple
generations (the complete algorithm is presented in Section \ref{sec:allocation}). 
Since each thread can now allocate memory in multiple generations,
multiple TLABs are necessary to avoid costly memory allocations. The TLAB to
use for each allocation is decided at runtime, based on the use of {\tt @Gen}
annotations (see Section \ref{sec:int_jit} for more details). Additionally to 
TLABs, \XPTO~also extends the use of ARs to multiple generations. Therefore,
whenever a TLAB used for a particular generation is full, an allocation request is
issued directly to the AR of the specific generation.

By using multiple TLABs and ARs (one for each generation),
allocations are more efficient as fewer synchronization barriers exist compared
to not using (TLABs and ARs). This, however, introduces a problem: 
as any thread can allocate memory in any generation, each thread must have a 
TLAB in each generation (even if that thread never allocates memory in that particular 
generation). As the number of generations grow, more and more memory is wasted for
allocating TLABs that are never actually used.

To solve the aforementioned problem, \XPTO~never actually allocates any memory
for TLABs when we create a new generation. Memory for each TLAB is 
effectively allocated only upon the first allocation request. This way, threads
will have TLABs (with allocated memory) only for the generations that are being 
used (and not for all the existing generations).

\subsection{{\tt @Gen} Annotations}
For allocating memory in generations other than \textit{Gen 0}, 
we considered several options: i) simply calling the JVM to switch the generation 
to use for allocation; ii) add a new {\tt new} operator with 
an extra argument (target generation); iii) annotate the {\tt new} operator.

We opted for the last option for the following reasons.
The first was immediately ruled out because it is very
difficult to control which objects go into non-young generations;
e.g., na{\"i}ve {\tt String} manipulation can easily result in many
allocations that would potentially go into a non-young generation. The second option
(creating a new allocation operator) would force us to extend the Java
language, and the compiler. 

A clear advantage of using annotations is its simplicity; however, 
it has one disadvantage: we must call the JVM whenever we need to 
change the current target generation. However, in practice and 
according to our experience, this almost never imposes a relevant 
overhead because:
i) a thread handling a particular task will most probably only need
one generation (worker threads tend to use one generation at
a time), and ii) large object allocation and copying is much more expensive than
calling the JVM to change the target generation (therefore it pays off
to allocate a large object in the correct generation). In both cases,
the cost of calling the JVM is absorbed and the overhead becomes
negligible 
(see Section~\ref{sec:evaluation} where we
show that \XPTO~does not decrease the application throughput). 
Also note that getting and setting the current generation
does not require any locking as it only changes a field in the current
thread's internal data structure.

\subsection{Code Interpreter and JIT}
\label{sec:int_jit}
The OpenJDK HotSpot uses a combination of code interpretation and Just-in-Time
(JIT) compilation to achieve close to native performance. Therefore, whenever
a method is executed for the first time, it is interpreted. If the
same method is executed for a specific number of times, it is then JIT compiled.
This way, the JVM compiles (a costly operation) only the methods where there is 
benefit (since executing compiled code is much faster than interpreting it).

In order to comply with such techniques in \XPTO, we modify both the
interpreter and the JIT compiler to add the notion of generations. To
be more precise, we had to detect if the allocation is annotated
with {\tt @Gen} and, if so, which generation is being targeted (choose
the correct TLAB).

Selecting the correct TLAB to allocate is done as follows. For each thread, 
\XPTO~keeps a pointer to the current generation TLAB. 
This pointer is only updated when the thread calls 
{\tt  newGeneration} or {\tt setGeneration}. Then, if the current
allocation site is annotated with {\tt @Gen}, the current generation
TLAB is used.

Detecting if the current allocation is annotated with {\tt @Gen} is
done differently before (interpretative mode) and after JIT compilation.
Before JIT, we use a map of byte code index to annotation, that is stored
along the method metadata (this map is prepared during class loading). 
Using this map, it is possible to know in constant time if a
particular byte code index is annotated with {\tt @Gen} or not. Upon
JIT compilation, the decision of whether to go for \textit{Gen 0} or not
is hardcoded into the compiled code. This frees the compiled code (after 
JIT) from accessing the annotation map.

\section{Evaluation}
\label{sec:evaluation}
We now evaluate the performance of \XPTO~while comparing it with G1, CMS. We also
have results for C4. Although not being an OpenJDK collector, C4 comes from a 
similar JVM, Zing\footnote{Zing is a JVM developed by Azul Systems (www.azul.com).}.
Since we only have on license available, we could not run all experiments with it. 

We use three relevant
platforms that are used in large-scale environments: 
i) Apache Cassandra 2.1.8 \cite{lakshman-2010}, a large-scale Key-Value store, 
ii) Apache Lucene 6.1.0 \cite{mccandless-2010}, a high performance text search engine, 
and iii) GraphChi 0.2.2 \cite{kyrola-2012}, a large-scale graph computation engine. 
A complete description of each workload, including how the source code was
changed (with the help of OLR profiler), is presented in Section \ref{sec:workloads}.
Please note that we could not use the widely studied DaCapo \cite{dacapo-2006}
benchmark suite since it does not compile for Java 8 (which is required for \XPTO~to
work). Nevertheless, we could run SPECjvm2008 \cite{shiv-2009} benchmarks whose 
results are published online at {\tt github.com/rodrigo-bruno/ng2c}.

For evaluating \XPTO, we are mostly concerned on showing that, compared with other
collectors, \XPTO: i) does reduce application pause times; ii) does not have
a negative effect on throughput nor for memory  utilization; iii) 
greatly reduces object copying; iv) does not increase the remembered set management work.

\subsection{Evaluation Setup}

\begin{table*}[!t]
\centering
\begin{tabular}{ c | c | c | c | c | c | c | c}       
Platform & Workload  & CPU & RAM & OS & Heap Size & Young/\textit{Gen 0} Size & LOC Changed \\
\hline
Cassandra & Feedzai  & Intel Xeon E5-2680 & 64 GB  & CentOS 6.7 & 30 GB & 4 GB & 22\\
\hline
Cassandra & WR,RW,RI & Intel Xeon E5505   & 16 GB  & Linux 3.13 & 12 GB & 2 GB & 22\\
\hline
Lucene    & RW       & AMD Opteron 6168   & 128 GB & Linux 3.16 & 120 GB & 2 GB & 8\\
\hline
GraphChi & PR,CC     & AMD Opteron 6168   & 128 GB & Linux 3.16 & 120 GB & 6 GB & 9\\
\end{tabular}       
\caption{Evaluation Environment Summary}   
\label{tab:eval_sum}
\end{table*}

We evaluate \XPTO~in three different environments (Table \ref{tab:eval_sum} provides a 
summary of the evaluation environments). First, we use Feedzai's\footnote{Feedzai
(www.feedzai.com) is a world leader data science company that detects fraud in
omnichannel commerce. The company uses near real-time machine learning to analyze big
data to identify fraudulent payment transactions and minimize risk in the financial
industry.} internal benchmark environment. This environment mirrors a real-world deployment and uses a Cassandra cluster to store data. For Feedzai, it is very
important to keep Cassandra's GC pauses as short as possible to guarantee that client
SLAs are not broken by long query lantencies. The Cassandra cluster is composed by 5
nodes.

Second, we use a separate node to evaluate \XPTO~with Cassandra under three different
synthetic workloads with varying number of read and write operations (more details in
Section \ref{sec:cass_workload}): Write-Intensive (WI), Write-Read (WR) and Read-
Intensive (RI).

Given the size of the data sets used for Lucene (Wikipedia dump) and GraphChi 
(Twitter graph dump), we use another separate node to evaluate \XPTO. On top
of Lucene we perform client searches while continuously updating the index
(read and write transactions). For GraphChi, we use two workloads, PageRank
and Connected Components. More details in Sections \ref{sec:lucene_workload} and
\ref{sec:graphchi_workload} (for Lucene and GraphChi workloads, respectively).

Each experiment runs in complete isolation for at least 5 times (i.e., until the
results obtained become stable). Feedzai's workload runs for 6 hours, while all other
workloads run for 30 minutes each. When running each experiment, we never consider the
first minute of execution (in Feedzai's benchmarks we disregard the first hour of
execution to allow other external systems to converge). This ensures minimal
interference from JVM loading, JIT compilation, etc. 

We always use fixed heap and young generation/\textit{Gen 0} sizes (see Table \ref{tab:eval_sum}). We
found that these sizes are enough to hold the working set in memory and to 
avoid premature massive promotion of objects to older generations (in the case of CMS and G1).
Table \ref{tab:eval_sum} also reports the number of lines changed after using the OLR
profiler.

\begin{figure*}[!t]
\centering
\begin{tabular}{cccc}
\subfloat[Cassandra WI]{\includegraphics[width=.23\textwidth]{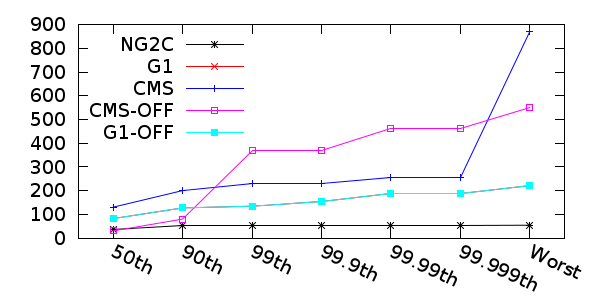}} &
\subfloat[Cassandra WR]{\includegraphics[width=.23\textwidth]{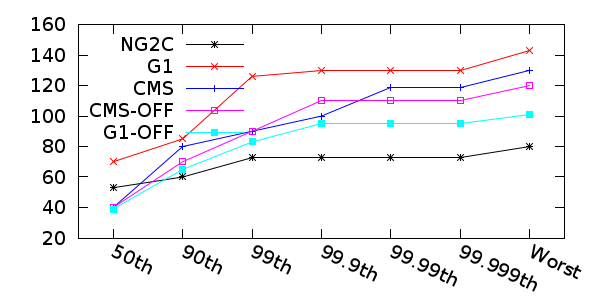}} &
\subfloat[Cassandra RI]{\includegraphics[width=.23\textwidth]{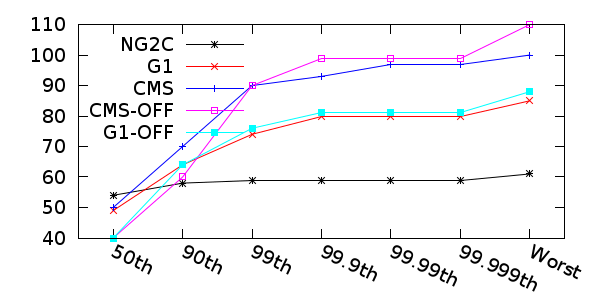}} &
\subfloat[Feedzai]{\includegraphics[width=.23\textwidth]{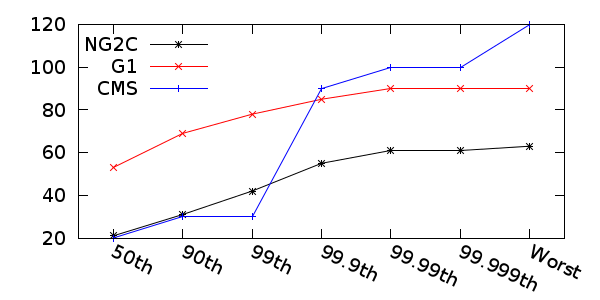}}
\end{tabular}
\begin{tabular}{ccc}
\subfloat[Lucene]{\includegraphics[width=.23\textwidth]{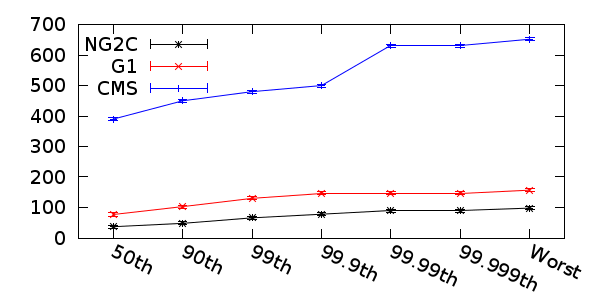}} &
\subfloat[GraphChi CC]{\includegraphics[width=.23\textwidth]{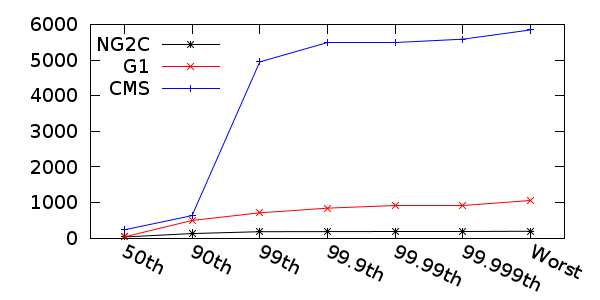}}&
\subfloat[GraphChi PR]{\includegraphics[width=.23\textwidth]{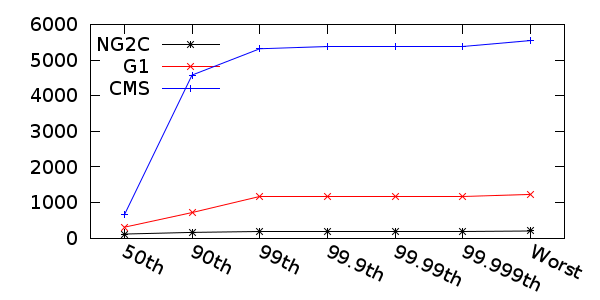}}
\end{tabular}
\caption{Pause Time Percentiles (ms)}
\label{fig:pause_percentiles}
\end{figure*}

\begin{figure*}[!t]
\centering
\begin{tabular}{cccc}
\subfloat[Cassandra WI]{\includegraphics[width=.23\textwidth]{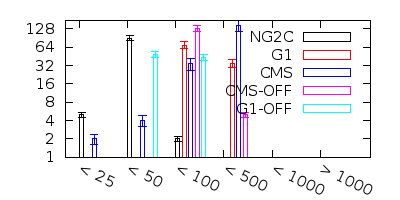}} & 
\subfloat[Cassandra WR]{\includegraphics[width=.23\textwidth]{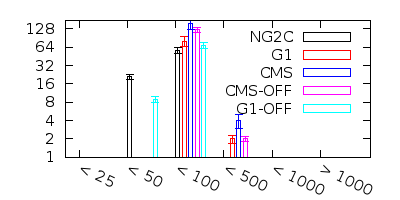}} &    
\subfloat[Cassandra RI]{\includegraphics[width=.23\textwidth]{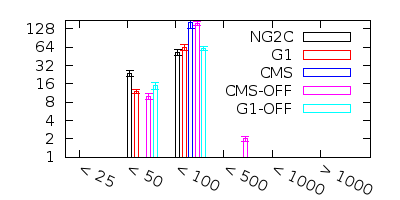}} &
\subfloat[Feedzai]{\includegraphics[width=.23\textwidth]{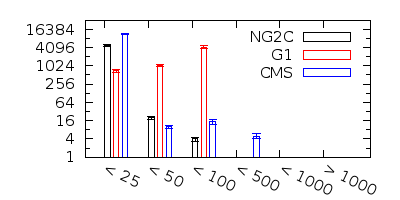}}
\end{tabular}
\begin{tabular}{ccc}
\subfloat[Lucene]{\includegraphics[width=.23\textwidth]{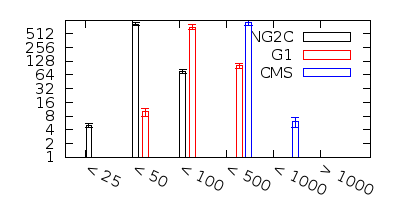}} &
\subfloat[GraphChi CC]{\includegraphics[width=.23\textwidth]{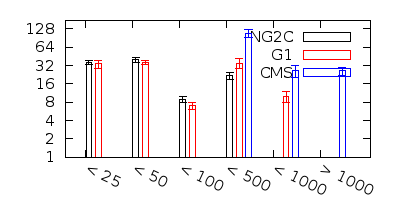}}&
\subfloat[GraphChi PR]{\includegraphics[width=.23\textwidth]{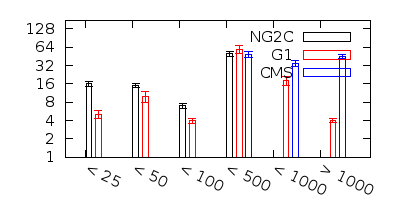}}
\end{tabular}
\caption{Number of Application Pauses Per Duration Interval (ms)}
\label{fig:pause_distribution}
\end{figure*}

\subsection{Workload Description}
\label{sec:workloads}
We use this section to provide a more complete description of the workloads used
to evaluate \XPTO.

\subsubsection{Cassandra}
\label{sec:cass_workload}
We use Cassandra under 4 different workloads: i) Feedzai's workload
(consisting of 500 read queries and 25000 write queries per second, for the whole
Cassandra cluster); ii) write
intensive workload (2500 read queries and 7500 write queries per second); iii)
read-write workload (5000 read queries and 5000 write queries per second); iv)
read intensive workload (7500 read queries and 2500 write queries per second).

Note that Feedzai's workload is based on representative data from real deployments
of their product (i.e., fraud detection). All workloads besides Feedzai's 
are synthetic but mirror real-world settings (e.g., we use the YCSB benchmark 
tool).\footnote{The Yahoo! Cloud Serving Benchmark (YCSB) is an 
open-source benchmarking tool often used to compare NoSQL database systems.}
When running Cassandra in Feedzai's cluster, we setup the JVM with 30GB of heap
and we fix the young generation (\textit{Gen 0} for \XPTO) to 4GB. 
Note that we tested several heap sizes and found these ones to be particularly 
good for short GC pause times.

To use \XPTO~we profiled Cassandra using the OLR profiler. The code was mainly
modified to allocate all objects
belonging to a particular {\tt Memtable}\footnote{A {\tt Memtable} table buffers recent
writes in memory. When a {\tt Memtable} is full, a flush is scheduled and a new
{\tt Memtable} is created. The capacity of each {\tt Memtable} is proportional
to the JVM heap size.} in a separate generation. Thus, whenever a new 
{\tt Memtable} is created or flushed, we create a new generation. 
Each {\tt Memtable} contains a B-Tree (self-balancing 
tree data structure) with millions of objects. These objects contain references 
to buffers with real data. To take advantage of \XPTO, we allocate all objects 
and buffers belonging to a particular {\tt Memtable} in the generation created
for that specific {\tt Memtable}.

In total, we changed a total of 22 code locations: i) 11 code locations where we 
annotate the {\tt new} operator, and ii) 11 code locations where we create, 
or change generation.

\subsubsection{Lucene}
\label{sec:lucene_workload}
We use Lucene to build an in-memory text index using a Wikipedia dump
from 2012.\footnote{Wikipedia dumps are available at dumps.wikimedia.org} The
dump has 31GB and is divided in 33M documents. Each document is loaded
into Lucene and can be searched. 

The workload is composed by 20000 writes (document updates) 
and 5000 reads (document searches) per second; note that this is a write intensive workload 
which represents a worst case scenario for GC pauses.
For reads (document queries), we loop through the 500 top words in the dump ; this also 
represents a worst case scenario for GC pauses.

When running Lucene, we use all available cores (48 cores), the heap size 
is limited to 120GB with a 2GB young generation (\textit{Gen 0} for \XPTO) size. Again, we tested 
with different heap sizes and we found out that this value is 
beneficial for short GC pauses.

To reduce Lucene's GC pauses
we profiled it with the OLR profiler. The code was mainly modified to allocate 
documents' data (of the Wikipedia dump) in a single separate generation. 
Objects created to hold the indexes of documents will live
throughout the application lifetime; therefore, if we do not use \XPTO~such objects
would be copied within the heap (thus leading to long GC pauses).
With \XPTO, most objects holding the index (including objects such as {\tt Term}, 
{\tt RAMFile} and {\tt byte} buffers) are allocated outside \textit{Gen 0}, 
i.e., in a separate generation. To accomplish it, we changed 8 code locations 
in Lucene, all of which to annotate the {\tt new} operator. 

\subsubsection{GraphChi}
\label{sec:graphchi_workload}
When compared to the previous systems (Cassandra and Lucene), GraphChi is a 
more throughput oriented system (and not latency oriented). 
However, we use GraphChi for two reasons: 
i) we want to demonstrate that \XPTO~does not decrease throughput even in 
a throughput oriented system; 
ii) with \XPTO, systems such as GraphChi can now be used for applications providing
latency oriented services, besides performing throughput oriented graph 
computations.

In our evaluation, we use two well-known algorithms: i) page rank, and ii) connected
components. Both algorithms are feed with a 2010 twitter graph \cite{kwak-2010}
consisting of 42 millions vertexes and 1.5 billions edges. 
These vertexes (and the corresponding
edges) are loaded in batches into memory; similarly to Cassandra' {\tt Memtables}, 
GraphChi calculates a memory budget to determine the number of edges to load into 
memory before the next batch. This represents an iterative process; in each iteration
a new batch of vertexes is loaded and processed.

When running GraphChi, we use all available cores (48 cores), the heap is
limited to 120GB, and the young generation (\textit{Gen 0} for \XPTO) is limited to 
6GB (we measured with different sizes and we found that this provides the shortest
GC pause times in the current environment and workload). 

To take advantage of \XPTO, we profiled GraphChi with the OLR profiler. The code
was mainly modified to allocate objects representing graph vertexes 
({\tt ChiVertex}), edges ({\tt Edge}), and internal pointers ({\tt ChiPointer}) in
multiple generations (one per batch). We modified a total of 9 code locations, in
which we annotate the {\tt new} operator.

\subsection{GC Pause Times}
Figure \ref{fig:pause_percentiles} presents the GC pause times for each GC (CMS, G1, 
and \XPTO) and for each percentile, for all the workloads. We do not show pause times
for C4 because it is a concurrent collector and therefore, the application should never
be paused. In practice, using C4, we got pauses of only up to 15 milliseconds for Cassandra.

In Feedzai's workload, GC pauses are shorter when compared to the
other Cassandra workloads. This is mainly because the hardware used in Feedzai 
achieves better performance compared to the one used for running the other 
Cassandra workloads. 
Still regarding Feedzai's workload, CMS shows shorter GC pauses for lower 
percentiles but shows the worst results in higher percentiles (25\% worse than
G1 and 47\% worse than \XPTO). G1 shows more stable GC pause times 
(when compared to CMS) as it does not lead to long pauses in higher percentiles; 
\XPTO~shows GC pause times very similar to CMS in lower percentiles, and it
shows shorter GC pause times for higher percentiles as well.

The other Cassandra workloads (WI, WR, and RI) differ only in the percentage of 
read and writes. From the GC perspective, more writes means that more objects are kept in memory 
(which results in more object copies and therefore longer GC pauses). 
This obviously applies to Cassandra because it buffers writes in memory.
This is clearly observable by comparing the GC pauses across the three 
workloads (WI, WR, and RI) for CMS and G1.
RI workload shows shorter GC pauses than WR and WI, while WR shows
shorter pauses than WI but longer than RI. According to our results,
CMS is more sensitive to writes (than the other two collectors) as it has a steep
increase in the GC pause time as we move towards write intensive workloads.
G1 has a more moderate increase in GC pause time in more intensive workloads.

Regarding \XPTO, it produces a different behavior as it shows shorter 
GC pauses for lower percentiles in WI, and longer pauses for WR in 
higher percentiles. One factor contributes for this difference 
(between \XPTO,~and G1 and CMS): we profiled (using OLR profiler) 
Cassandra under the WI workload. This means that the read path is not 
as optimized as the write path. Therefore, in write intensive workloads, \XPTO~is 
more optimized than in read intensive workloads. This is also observable by 
measuring the difference between the GC pause times in higher percentiles;
as we move towards write intensive workloads, the difference between 
\XPTO~and other GCs increases.

We also have results for Cassandra with the off-heap memory enabled for CMS and
G1 (Cassandra uses off-heap memory to store values while the keys remain in the 
managed heap). Using off-heap reduces GC pause times by up to 50\% in the WI workload 
(versus 93.8\% using \XPTO), around 20\% in the WR workload (versus 39\% using \XPTO), 
and shows no improvement for the RI workload (versus 61\% using \XPTO). In sum, 
using \XPTO~is more effective to reduce GC pause times than using off-heap memory
mainly because Cassandra needs to keep header objects in the memory managed heap
to describe the contents stored in off-heap. In the case of Cassandra (key-value
store), keys are stored in the managed heap and therefore contribute for long
application pauses. \XPTO~is able to move all key-value pairs into a specific
generation (thus avoiding pause times).

The remaining workloads (Lucene, PR, and CC) are all write intensive. CMS shows 
very high GC pause times compared to the other two GCs. G1 shows a more
moderate increase in GC pause times, when compared to CMS, but is still worse 
than \XPTO. In sum, \XPTO~clearly improves the worst observable GC pause times by: 
85\% (CMS) and 38\% (G1) in Lucene, 
97\% (CMS) and 84\% (G1) in PR, and 
97\% (CMS) and 82\% (G1) for CC.

Figure \ref{fig:pause_distribution} presents the average and standard deviation for the
number of pauses in different
duration intervals. Results show that: i) \XPTO~does not increase the number
of pauses, and ii) it moves pauses to smaller duration intervals. CMS presents
the worst results by having the most amount of pauses in longer pause intervals.


\subsection{Object Copy and Remembered Set Update}
\begin{figure}[!t]
\centering
\begin{tabular}{cc}
\subfloat[Object Copy Norm. to G1]{\includegraphics[width=.23\textwidth]{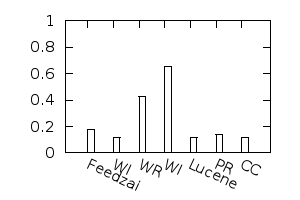}} & 
\subfloat[Rem. Set Update Norm. to G1]{\includegraphics[width=.23\textwidth]{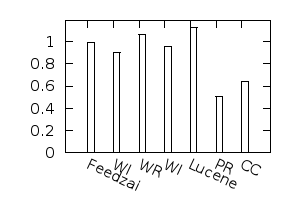}}
\end{tabular}
\caption{\XPTO~Object Copy and Remembered Set Update}
\label{fig:copy-red}
\end{figure}

\begin{figure*}[!t]
\centering
\begin{tabular}{ccc}
\subfloat[Cassandra WI]{\includegraphics[width=.32\textwidth]{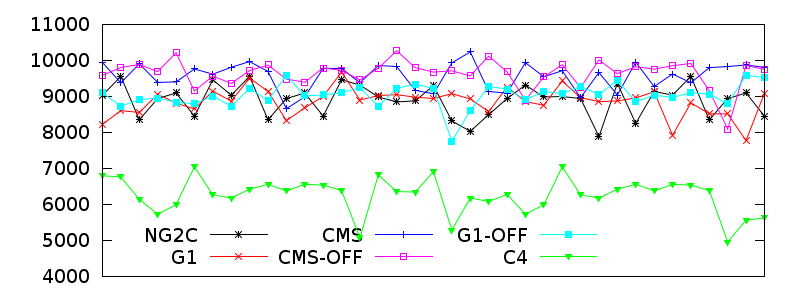}} & 
\subfloat[Cassandra WR]{\includegraphics[width=.32\textwidth]{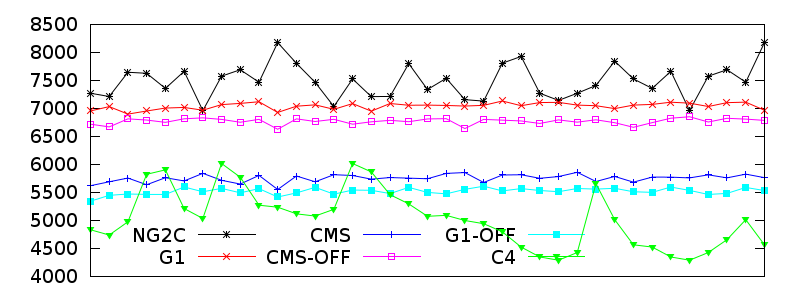}} & 
\subfloat[Cassandra RI]{\includegraphics[width=.32\textwidth]{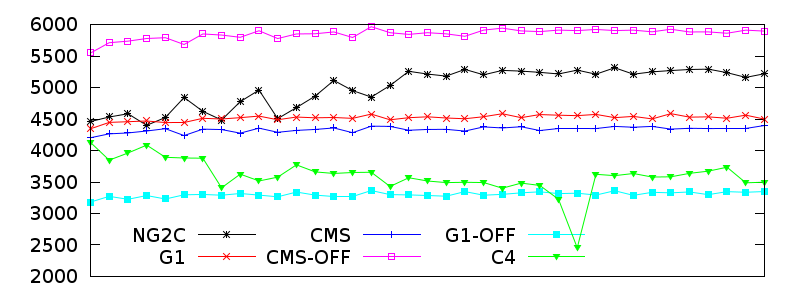}}
\end{tabular}
\caption{Cassandra Throughput (transactions/second) - 10 minutes sample}
\label{fig:cassandra_throughput}
\end{figure*}

We now look into how much time is spent: i) copying objects within the heap, and ii) 
updating remembered set entries, upon a collection. Note that the remembered set updates
is an important metric since pretenuring can lead to high number of remembered set updates 
because of the potential increase in the number of references coming from older to younger
spaces \cite{jones-2016}. We only show results for G1 and 
\XPTO, given that CMS and C4 do not provide such logging information. However, both metrics
are similar for different generational collectors because they mostly depend on: 
i) the mutator allocation 
speed (dictates how fast minor collections are triggered and how many objects are
promoted), and ii) the available hardware memory bandwidth. Both these factors are
kept constant across GCs (G1 and \XPTO).

Figure \ref{fig:copy-red} presents results for total object copying time and remembered set
update time during each workload. All results are normalized to G1. Results show that
\XPTO~reduces objects copying between 30.6\% and 89.2\%. Note that, in G1, we can not
differentiate between object promotion and object compaction since the collector
collects both young and old regions at the same time (during mixed collections).

\XPTO~also has a positive impact for the remembered set update work. This means that,
in \XPTO, there is not an increase in the number of references from older generations into
\textit{Gen 0}. This is possible because objects referenced by pretenured objects are most 
likely to be pretenured as well. \XPTO~even reduces the amount of remembered set update work for
most workloads since it reduces the amount of premature promotion in G1 (objects with short
lifetimes that were allocated right before a minor collection and were prematurely promoted).
This also means that \XPTO~puts less pressure on the write barrier (compared to G1).

\begin{table}[!t]
\centering
\begin{tabular}{ l | c | c | c | c | c | c}       
 & \multicolumn{3}{c|}{Max Mem Usage} & \multicolumn{3}{c}{Throughput}\\
\hline
 & CMS & G1 & C4 & CMS/OFF& G1/OFF & C4\\
\hline
Feedzai  & .92 & 1.00 & - & - & - & -\\
WI       & .96 & 1.01 & 1.73 & 1.07/1.08 & .99/1.01 & .70\\
WR       & .80 & 1.00 & 2.04 & .76/.90 & .93/0.73 & .67\\
RI       & .73 & .98 & 1.94 & .86/1.18 & .90/0.65 & .71\\
\hline
Lucene   & .39 & .98 & - & .59 & .87 & -\\
\hline
PR       & 1.44 & 1.04 & - & .80 & .96 & -\\
CC       & 1.43 & 1.17 & - & 1.03 & .96 & -\\
\end{tabular}       
\caption{Max Memory Usage and Throughput norm. to \XPTO~(i.e., \XPTO~value is 1 for all
entries)}  
\label{tab:heap}
\end{table}

\subsection{Memory Usage}
In this section, we look into the max memory usage to understand how \XPTO~relates
to other collectors regarding heap requirements (see Table \ref{tab:heap}). 
Regarding the workloads' max heap size: Feedzai workload has 30GB, while the other
Cassandra workloads (WI, WR, and RI) have 12GB; each Lucene and GraphChi's
workload (PR and CC) have 120GB.

From Table \ref{tab:heap} we can conclude that, regarding Cassandra workloads 
(i.e., Feedzai, WI, WR, and RI) all collectors (excluding C4) have a very similar max 
memory usage. CMS has a slight smaller heap (compared to G1 and \XPTO) while 
\XPTO~has a slight larger heap (compared to G1 and CMS).
This slight increase comes from the fact that generations are only collected
upon a mixed collection, which is only triggered when the heap usage is above
a configurable threshold. This can lead to a slight delay in the collection of 
some objects that are already unreachable. C4 has a considerable higher memory
usage since it reserves approximately 75\% (12GB) of the system's memory, when
the JVM is launched. We were unable to extract the actual memory usage during
execution. We do not show the results for C4 with other workloads because we 
only have one license (for one physical node).

Lucene max memory utilization is lower for CMS when compared to G1 and \XPTO. 
These larger heap sizes in G1 and \XPTO~comes from humongous allocations. 
Using this technique, very large objects are directly allocated in the
old generation. It has the clear drawback of delaying the collection of such very 
large objects. Since
CMS does not have such technique (i.e., all objects are allocated in the {\tt Eden}),
CMS tries to collect these large objects upon each minor collection, leading to
faster collection of such objects, thus achieving lower heap usage. 
Comparing G1 with \XPTO, the heap usage is similar.
 
Regarding GraphChi, which shows a different memory behavior when compared to  
Cassandra and Lucene, it allocates many small objects (which therefore are not 
considered for humongous allocation in G1 and \XPTO). Most of these small objects 
(mostly data objects representing vertexes and edges) are used in a single iteration, 
which is long enough for them to be promoted into the old generation (in the case of 
CMS and G1). Since we set the maximum heap size to 120GB, 
the heap fills up until a concurrent marking cycle is triggered. 
In CMS, the concurrent marking cycle is triggered a bit later compared to G1 and 
\XPTO~(thus leading to an increase in the max heap usage). Regarding G1 and \XPTO, 
both present similar max heap values.

\subsection{Application Throughput}
\label{sec:throughput}

We now discuss the throughput obtained for each GC and workload 
(except for Feedzai). We do not show the throughput for Feedzai's workload 
because the benchmark environment (where the Cassandra cluster is used) 
dynamically adjusts the number of transactions per second 
according to external factors; e.g., the credit-card transaction generator produces 
different transactions through time, some result on more Cassandra transactions 
than others, thus making it infeasible to reproduce the same workload multiple times.
The throughput for all remaining workloads is presented in Table \ref{tab:heap}.
Throughput for Cassandra using off-heap is shown for WI, WR, and RI workloads.
All results are normalized to \XPTO.

From Table \ref{tab:heap}, we conclude that \XPTO~outperforms
CMS, G1, and C4 (we could only obtain results for Cassandra workloads using
C4 because we only have one license) for most workloads. Figure \ref{fig:cassandra_throughput} 
shows the throughput evolution for Cassandra workloads. \XPTO~is the solution with overall best 
throughput across the three workloads. Only the CMS collector using off-heap outperforms 
\XPTO~in the read intensive workload (by approximately 18\%).


For all previous experiments, we use latency oriented GC configurations,
i.e., the configurations we found to enable shorter GC pause times in higher 
percentiles. This, however, has the drawback of potentially decreasing the 
throughput. Among the used workloads, the most explicit example of this throughput
decrease is Lucene running with CMS, in which a throughput oriented GC configuration,
i.e., the configuration we found to enable higher throughput,
could increase the throughput by up to 3x (when compared to the throughput achieved
with a latency oriented configuration).

\begin{figure}[t]                                                        
\centering                                                                   
\includegraphics[keepaspectratio,width=.4\textwidth]{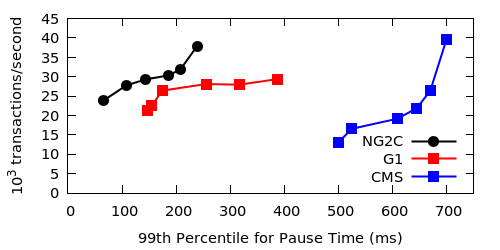}
\caption{Throughput vs Pause Time}
\label{fig:throughput_vs_latency}
\end{figure}

To better understand the trade-off between throughput and latency, we ran the 
Lucene workload with 6 different young generation (G1 and CMS) or \textit{Gen 0} (\XPTO) sizes. 
We found that this parameter alone allows one to achieve good latency (if the size is reduced) 
or good throughput (if the size is increased). Other GC parameters did not have a relevant effect 
and therefore we keep them fixed. We start with the configuration used in the previous sections
(2 GB). Then, we keep increasing the size of the young generation/\textit{Gen 0} by 2 GB.

Figure \ref{fig:throughput_vs_latency} 
shows a plot with the relation between throughput and GC pause time, for each GC, in which
each point on each line represents a different young generation/\textit{Gen 0} size.
CMS shows always longer GC pauses independently of the GC configuration. 
It also shows a steep increase in the throughput, with a small increase in 
the GC pause time; this shows how easy it is to dramatically reduce throughput 
when CMS is configured for latency.
On the other hand, G1 shows much shorter GC pauses than CMS at the cost of some 
reduced throughput. Note that moving from latency oriented to throughput oriented 
configurations has a small impact on throughput, but has a larger negative impact on 
GC pause time. 
Finally, \XPTO~provides the shortest GC pause times with a very small throughput
impact. In the most throughput oriented configuration (point on the top of the curve), 
\XPTO~is only 5\% worse than CMS and the GC pause time is 66\% better. 
In conclusion: 
i) CMS can be difficult to configure for short GC pause time (while keeping an acceptable 
throughput); 
ii) G1 leads to shorter pauses but can damage throughput; 
iii) \XPTO~keeps up with the best throughput achieved by CMS, while also 
reducing the GC pause times by 66\% and 39\% w.r.t. CMS and G1, respectively.

\section{Conclusions and Future Work}
This paper presents the design and implementation of \XPTO,\footnote{Both \XPTO~and
the OLR profiler can be 
downloaded from {\tt github.com/rodrigo-bruno/ng2c}} a new HotSpot GC algorithm that avoids 
copying objects within the heap by aggregating objects with similar lifetime
profiles in separate generations. \XPTO~is built on top of G1, by modifying the way
it allocates objects and manages generations.
The experimental evaluation shows that it is possible 
to reduce the object copying done by current 
collectors (G1 and CMS) by up to 89.2\%, resulting in shorter GC pause times. We are able to 
reduce the worst observable GC pause times in Cassandra by 94.8\%, 85\% for Lucene, and 
96.45\% for GraphChi. We also show that despite increasing the complexity of 
the JVM allocation algorithm, \XPTO~does not penalize application throughput, the heap
usage, and the remembered set update work, when compared to current GC implementations.

We envision that the \XPTO~could be integrated in other JVMs and collectors. Even
concurrent collectors such as C4 could take advantage of the ideas described in this
work to reduce the amount of object copying within the heap and therefore reduce the
application interference, possibly increasing the throughput. We
are currently working on integrating the object lifetime estimation directly into
the JVM in order to allow dynamic N-Generational pretenuring.
                                               

\bibliographystyle{abbrvnat}
\bibliography{main}

\end{document}